\documentclass[11pt,a4paper]{article}
\pdfoutput=1
\usepackage{jheppub}
\usepackage{tikz}
\usetikzlibrary{intersections, calc, arrows.meta, bending}
\usepackage{bbm}

\DeclareMathOperator{\Tr}{Tr}
\DeclareMathOperator{\tr}{tr}

\newcommand{\ri}{\mathrm{i}}
\renewcommand{\th}{\theta}

\newcommand{\cob}{\delta}

\newcommand{\hf}{\frac{1}{2}}
\newcommand{\qu}{\frac{1}{4}}
\newcommand{\til}[1]{\widetilde{#1}}

\renewcommand{\b}[1]{\overline{#1}}

\newcommand{\bra}{\langle}
\newcommand{\ket}{\rangle}
\newcommand{\la}{\lambda}

\newcommand{\bt}{\beta}
\newcommand{\ga}{\gamma}

\newcommand{\al}{\alpha}
\newcommand{\om}{\omega}

\newcommand{\rt}[1]{\sqrt{#1}}
\newcommand{\cO}{\mathcal{O}}
\newcommand{\cZ}{\mathcal{Z}}

\newcommand{\id}{\mathbbm{1}}
\makeatletter
\gdef\@fpheader{}
\makeatother

\begin{document}
\title{Cap amplitudes in random matrix models}

\author{Kazumi Okuyama}

\affiliation{Department of Physics, 
Shinshu University, 3-1-1 Asahi, Matsumoto 390-8621, Japan}

\emailAdd{kazumi@azusa.shinshu-u.ac.jp}

\abstract{
For general one-matrix models in the large $N$ limit,
we introduce the cap amplitude $\psi(b)$ 
as the expansion coefficient of the 1-form $ydx$
on the spectral curve.
We find that the dilaton equation for the
discrete volume $N_{g,n}$ of the moduli space
of genus-$g$ Riemann surfaces with $n$ boundaries
is interpreted as gluing the cap amplitude along one of the boundaries.
In this process, one of the boundaries is capped 
and the number of boundaries decreases by one.
In a similar manner, the genus-$g$ free energy $F_g$ is obtained
by gluing the cap amplitude to $N_{g,1}$.
}

\maketitle

\section{Introduction}\label{sec:intro}
Random matrix models are ubiquitous tools in physics and mathematics.
In particular, the double scaling limit of
matrix models plays an important role
in the study of 2d quantum gravity \cite{Brezin:1990rb,Douglas:1989ve,Gross:1989vs}.
Recently, in the 
seminal paper \cite{Saad:2019lba}
it was found that JT gravity is described by
a double-scaled matrix model as well, which opened a new avenue for applying
the matrix model technology to study the holography in quantum gravity.
As shown in \cite{Saad:2019lba}, the correlator of the partition function
$Z(\bt)$ of JT gravity is constructed by
gluing some basic building blocks, 
the trumpet and the Weil-Petersson volume $V_{g,n}(b_1,\cdots,b_n)$
of the moduli space of genus-$g$ Riemann surfaces with $n$ boundaries
of length $b_i\in\mathbb{R}_{+}~(i=1,\cdots,n)$;
the correlator of $Z(\bt)$ is then obtained by integrating over the lengths $b_i$
of $n$ boundaries.

Interestingly, 
a similar story holds for the so-called ETH matrix model for DSSYK
\footnote{
ETH and DSSYK stand for
the eigenstate thermalization hypothesis and the double-scaled 
Sachdev-Ye-Kitaev model, 
respectively.} \cite{Jafferis:2022wez}. 
In \cite{Okuyama:2023kdo}, it was found that
the correlator of the ETH matrix model
is constructed by
gluing the trumpet and the discrete volume 
$N_{g,n}(b_1,\cdots,b_n)$
introduced in \cite{norbury2013polynomials}.
One important difference from the JT gravity case is that 
the boundary lengths $b_i$ in the discrete volume $N_{g,n}(b_1,\cdots,b_n)$ take
discrete values $b_i\in\mathbb{Z}_{+}$. 
Another difference from the JT gravity case is that 
the ETH matrix model is defined by the ordinary large $N$ limit
without taking the double scaling limit. 
Although we are not taking the double scaling limit,
the ETH matrix model is expected to be holographically dual
to some 2d quantum gravity;
one candidate for the bulk dual is the sine dilaton gravity \cite{Blommaert:2024ymv,Blommaert:2024whf}.

In \cite{Jafferis:2022wez,Okuyama:2024eyf}, it was realized that
the potential and the eigenvalue density of the ETH matrix model
are naturally expanded in the basis of Chebyshev polynomials 
$T_b(x)$. 
In \cite{Blommaert:2025rgw}, it was argued that the coefficient $\psi(b)$
of this expansion determines
the genus-zero free energy (or sphere amplitude)
of the ETH matrix model for DSSYK.
It turns out that the expansion of the potential in the basis of $T_b(x)$
is not a special feature of the ETH matrix model.
In fact, 
this expansion
has been considered in \cite{Borot:2010tr}
for the general one-matrix model with an arbitrary potential.
However, the geometric meaning of the coefficient $\psi(b)$
of this expansion has not been fully explored in the literature before, 
as far as we know.

In this paper, we call $\psi(b)$ the ``cap amplitude'', which is schematically 
depicted as \footnote{Cap amplitudes also appear 
in 2d Yang-Mills theory \cite{Cordes:1994fc,Aganagic:2004js}
and 2d TQFT (see e.g. \cite{kock2004frobenius})
and play important roles in these theories.}
\begin{equation}
\begin{aligned}
 \begin{tikzpicture}[scale=1]
\draw (1,-0.5) node [below]{$b$};
\draw (1,0.5) .. controls (-0.5,0.3) and (-0.5,-0.3) .. (1,-0.5);
\draw[thick,red] (1.25,0) arc [start angle=360,end angle=0, x radius=0.25, y radius=0.5];
\draw (-0.5,0) node [left]{$\psi(b)=$};
\end{tikzpicture}
\end{aligned} 
\label{fig:psi}
\end{equation}
We interpret $b\in\mathbb{Z}_{+}$ as the discrete length of the boundary of the cap.
Our geometric picture \eqref{fig:psi} of the cap amplitude 
comes from the dilaton equation obeyed by the discrete volume 
$N_{g,n}$.
In \cite{norbury2013polynomials}, it was proved that
$N_{g,n}$ is obtained by acting an operator on $N_{g,n+1}$,
where the operator is constructed from the data of the spectral curve
of the matrix model.
We find that the dilaton equation in \cite{norbury2013polynomials}
has a simple expression in terms of the 
cap amplitude $\psi(b)$ defined
by the expansion coefficient of the 1-form $ydx$ on 
the spectral curve. 
Our result of the dilaton equation reads
\begin{equation}
\sum_{b=0}^\infty
\psi(b)N_{g,n+1}(b,b_1,\cdots,b_n)
=(2-2g-n)N_{g,n}(b_1,\cdots,b_n).
\label{eq:dilaton}
\end{equation}
This is schematically depicted as
\begin{equation}
\begin{aligned}
\begin{tikzpicture}[scale=0.65]
\draw[thick,red] (1,0.5) arc [start angle=90,end angle=270, x radius=0.25, y radius=0.5];
\draw[thick,dashed,red] (1,0.5) arc [start angle=90,end angle=-90, x radius=0.25, y radius=0.5];
%\draw[thick,red] (1.25,0) arc [start angle=360,end angle=0, x radius=0.25, y radius=0.5];
\draw (1,-0.5) node [below]{$b$};
\draw (6.75,1.5) arc [start angle=360,end angle=0, x radius=0.25, y radius=0.5];
\draw (6.75,-1.5) arc [start angle=360,end angle=0, x radius=0.25, y radius=0.5];
%\draw (6.75,0) arc [start angle=360,end angle=0, x radius=0.25, y radius=0.5];
\draw (6.5,1) .. controls (5.8,0.95) and (5.8,0.55) .. (6.5,0.5);
\draw (6.5,-0.5) .. controls (5.8,-0.55) and (5.8,-0.95) .. (6.5,-1);
\draw (6.3,0.1) node [right] {$\vdots$};
\draw (6.8,1.5) node [right] {$1$};
\draw (6.8,-1.5) node [right] {$n$};
\draw (1,0.5) .. controls (-0.5,0.3) and (-0.5,-0.3) .. (1,-0.5);
\draw (1,0.5) .. controls (2.5,0.6) and (5,2) .. (6.5,2);
\draw (1,-0.5) .. controls (2.5,-0.6) and (5,-2) .. (6.5,-2);
\draw (2,0.1) to [out=-30,in=-150] (3,0.1);
\draw (2.2,0) to [out=30,in=150] (2.8,0);
\draw (4.5,0.1) to [out=-30,in=-150] (5.5,0.1);
\draw (4.7,0) to [out=30,in=150] (5.3,0);
\draw (3.8,-0.3) node [above] {$\cdots$};
\draw (2.5,-0.1) node [below] {$1$};
\draw (5,-0.1) node [below] {$g$};
\draw (-0.2,0) node [left]{$\displaystyle 
\sum_{b=0}^\infty$};
\draw (7.8,0) node [right]{$~=~(2-2g-n)$};
\draw (18,1) .. controls (17.3,0.95) and (17.3,0.55) .. (18,0.5);
\draw (18,-0.5) .. controls (17.3,-0.55) and (17.3,-0.95) .. (18,-1);
\draw (17.8,0.1) node [right] {$\vdots$};
\draw (18.3,1.5) node [right] {$1$};
\draw (18.3,-1.5) node [right] {$n$};
\draw (18.25,1.5) arc [start angle=360,end angle=0, x radius=0.25, y radius=0.5];
\draw (18.25,-1.5) arc [start angle=360,end angle=0, x radius=0.25, y radius=0.5];
\draw (13,0.5) .. controls (14,0.7) and (16,2) .. (18,2);
\draw (13,-0.5) .. controls (14,-0.7) and (16,-2) .. (18,-2);
\draw (13,0.5) .. controls (12.3,0.3) and (12.3,-0.3) .. (13,-0.5);
\draw (13.5,0.1) to [out=-30,in=-150] (14.5,0.1);
\draw (13.7,0) to [out=30,in=150] (14.3,0);
\draw (16,0.1) to [out=-30,in=-150] (17,0.1);
\draw (16.2,0) to [out=30,in=150] (16.8,0);
\draw (15.3,-0.3) node [above] {$\cdots$};
\draw (14,-0.1) node [below] {$1$};
\draw (16.5,-0.1) node [below] {$g$};
\end{tikzpicture}
\end{aligned}
\label{fig:dilaton}
\end{equation}
Namely, by gluing the cap amplitude $\psi(b)$
along one of the boundaries, that boundary is capped and becomes closed.
In this process,
the number of
boundaries decreases by one: $N_{g,n+1}\to N_{g,n}$.
Note that \eqref{eq:dilaton} holds when $(g,n)$ satisfies the condition
$2-2g-n<0$.
As a special case $n=0$ of \eqref{eq:dilaton}, we find the 
dilaton equation for the genus-$g$ free energy $F_g~(g\geq2)$
\begin{equation}
 F_g=\frac{1}{2-2g}\sum_{b=0}^\infty\psi(b)N_{g,1}(b).
\label{eq:conj}
\end{equation}
Again, \eqref{eq:conj} is schematically depicted as
\begin{equation}
\begin{aligned}
\begin{tikzpicture}[scale=0.8]
\draw[thick,red] (1,0.5) arc [start angle=90,end angle=270, x radius=0.25, y radius=0.5];
\draw[thick,dashed,red] (1,0.5) arc [start angle=90,end angle=-90, x radius=0.25, y radius=0.5];
%\draw[thick,red] (1.25,0) arc [start angle=360,end angle=0, x radius=0.25, y radius=0.5];
\draw (1,-0.5) node [below]{$b$};
\draw (1,0.5) .. controls (-0.5,0.3) and (-0.5,-0.3) .. (1,-0.5);
\draw (1,0.5) to [out=10,in=180] (4,1.6);
\draw (4,1.6) to [out=0,in=90] (6,0);
\draw (1,-0.5) to [out=-10,in=180] (4,-1.6);
\draw (4,-1.6) to [out=0,in=-90] (6,0);
\draw (2,0.1) to [out=-30,in=-150] (3,0.1);
\draw (2.2,0) to [out=30,in=150] (2.8,0);
\draw (4.5,0.1) to [out=-30,in=-150] (5.5,0.1);
\draw (4.7,0) to [out=30,in=150] (5.3,0);
\draw (3.8,-0.3) node [above] {$\cdots$};
\draw (2.5,-0.1) node [below] {$1$};
\draw (5,-0.1) node [below] {$g$};
\draw (-0.2,0) node [left]{$\displaystyle F_g=\frac{1}{2-2g}
\sum_{b=0}^\infty$};
\end{tikzpicture}
\end{aligned}
\label{fig:free}
\end{equation}

Our definition of $\psi(b)$ as the coefficient of $ydx$ is essentially the same as 
the expansion coefficient of the potential in the basis of 
Chebyshev polynomials considered in \cite{Borot:2010tr}. However, we stress that
the sum in \eqref{eq:dilaton} starts from $b=0$; in 
our definition the value of $\psi(0)$
is naturally
fixed, while the expansion of the potential
does not contain the information of $\psi(0)$.
This is the advantage of using the 1-form $ydx$ to define the cap amplitude
$\psi(b)$.

For general hermitian one-matrix models,
the technique to compute the genus expansion of the free energy has been
developed in \cite{Ambjorn:1992gw}. 
In general, $F_g$ is expressed as some combination of the moments $M_k,J_k$ 
\cite{Ambjorn:1992gw}.
In section \ref{sec:cap}, 
we will check that the genus-two free energy
$F_{2}$ computed from the dilaton equation \eqref{eq:conj}
agrees with the known result in \cite{Ambjorn:1992gw}.
Interestingly, we find that the moments $M_k,J_k$ themselves are written
as some combination of the cap amplitudes.
This implies that $F_g$ is completely determined by the 
information of the cap amplitude alone.

This paper is organized as follows.
In section \ref{sec:top}, we briefly review the topological recursion
for the hermitian one-matrix model in the one-cut case.
We also explain the dilaton equation for $\om_{g,n}$, 
the genus-$g$ correlator of $n$ resolvents.
In section \ref{sec:cap}, we introduce the cap amplitude 
$\psi(b)$ as the expansion coefficient of the 1-form $ydx$
on the spectral curve.
Following \cite{norbury2013polynomials},
we prove the dilaton equation \eqref{eq:dilaton}
by the partial integration and the contour deformation.
In the rest of section \ref{sec:cap}, we compare our 
approach with the existing literature \cite{Borot:2010tr} and
\cite{Ambjorn:1992gw}.
In section \ref{sec:example},
we consider the Gaussian matrix model and the ETH matrix model for DSSYK
as examples of our formalism.
We compute the cap amplitudes for these models and confirm the
dilaton equation for the free energy \eqref{eq:conj}.
Finally, we conclude in section \ref{sec:discussion} with some discussions on 
future problems. 

\section{Review of topological recursion}
\label{sec:top}

In this section, we will briefly review the topological recursion of 
the matrix model in the large $N$ limit.
We consider the hermitian one-matrix model
\begin{equation}
\begin{aligned}
 \cZ=\int dM e^{-N\Tr V(M)},
\end{aligned} 
\label{eq:hermitian-mat}
\end{equation}
where $M$ is an $N\times N$ hermitian matrix.
In the large 
$N$ limit, 
the free energy $\log \cZ$ admits the genus expansion 
\begin{equation}
\begin{aligned}
 \log\cZ=\sum_{g=0}^\infty N^{2-2g}F_g.
\end{aligned} 
\end{equation}

We are also interested in the correlator of the 
resolvent $\Tr\frac{1}{x-M}$.
It is known that the genus-zero resolvent
\begin{equation}
\begin{aligned}
 R(x)=\left\bra\frac{1}{N}\Tr\frac{1}{x-M}\right\ket_{g=0}
\end{aligned} 
\label{eq:Rx}
\end{equation}
satisfies the loop equation (see e.g. \cite{Eynard:2015aea} for a review)
\begin{equation}
\begin{aligned}
 R(x)^2-V'(x)R(x)+P(x)=0,
\end{aligned} 
\label{eq:loop}
\end{equation}
where $P(x)$ is given by
\begin{equation}
\begin{aligned}
 P(x)=\left\bra\frac{1}{N}\Tr\frac{V'(x)-V'(M)}{x-M}\right\ket_{g=0}.
\end{aligned} 
\end{equation}
From the loop equation \eqref{eq:loop},
we can define the spectral curve for the matrix model
\begin{equation}
\begin{aligned}
 y^2=\qu V'(x)^2-P(x),
\end{aligned} 
\label{eq:curve}
\end{equation}
where $y$ is given by
\begin{equation}
\begin{aligned}
 y=\hf V'(x)-R(x).
\end{aligned} 
\label{eq:y-def}
\end{equation}
The spectral curve \eqref{eq:curve} 
is a double cover of the $x$-plane with some number of branch points.
In this paper, we consider the so-called one-cut case,
where $y$ has two branch points
$x=a,b$ with a square-root branch cut $x\in[a,b]$
\begin{equation}
\begin{aligned}
 y=\hf Q(x)\rt{(x-a)(x-b)}.
\end{aligned} 
\label{eq:y-Q}
\end{equation}
Here $Q(x)$ is some regular function of $x$. 
In the one-cut case, it is convenient to
introduce the Joukowsky map
\begin{equation}
\begin{aligned}
 x(z)=\ga(z+z^{-1})+\cob.
\end{aligned} 
\label{eq:Joukowsky}
\end{equation}
In the $z$-coordinate, 
the branch points correspond to $z=\pm 1$
where $dx(z)$ vanishes.
From the condition
\begin{equation}
\begin{aligned}
 x(1)=a,\quad x(-1)=b,
\end{aligned} 
\end{equation}
 $\ga$ and $\cob$ in \eqref{eq:Joukowsky} are determined as
\begin{equation}
\begin{aligned}
 \ga=\frac{a-b}{4},\quad \cob=\frac{a+b}{2}.
\end{aligned} 
\label{eq:ga-cob}
\end{equation}
It turns out that the Joukowsky map \eqref{eq:Joukowsky}
defines a uniformization coordinate $z$ for the spectral
curve \eqref{eq:y-Q}.
Using the relation
\begin{equation}
\begin{aligned}
x-a&=\ga(z+z^{-1}-2),\\
x-b&=\ga(z+z^{-1}+2),
\end{aligned} 
\label{eq:xa-xb}
\end{equation}
the square-root in \eqref{eq:y-Q} is simplified as
\begin{equation}
\begin{aligned}
\rt{(x-a)(x-b)} =\ga(z-z^{-1}).
\end{aligned} 
\label{eq:root-z}
\end{equation}
One can see that the two branches of the square-root in \eqref{eq:y-Q}
are exchanged by sending $z$ to its conjugate point $\b{z}=z^{-1}$,
while $x(z)$ is invariant under $z\to\b{z}$
\begin{equation}
\begin{aligned}
 y(\b{z})=-y(z),\quad x(\b{z})=x(z).
\end{aligned} 
\end{equation}
By the Joukowsky map \eqref{eq:Joukowsky}, the cut $[a,b]$ on the $x$-plane
is mapped to the unit circle $|z|=1$ on the $z$-plane, and
the two sheets of $y$ are mapped to the inside $|z|<1$ or outside $|z|>1$
of the unit circle.

Next, let us consider the connected correlator of the resolvents.
We define the degree $n$ symmetric differential
$\om_{g,n}(z_1,\cdots,z_n)$ on the spectral curve by 
\begin{equation}
\begin{aligned}
 \left\bra\prod_{i=1}^n\Tr\frac{dx(z_i)}{x(z_i)-M}\right\ket_{\text{conn}}
=\sum_{g=0}^\infty N^{2-2g-n}\om_{g,n}(z_1,\cdots,z_n).
\end{aligned} 
\end{equation}
$\om_{g,n}$ can be systematically
computed 
by the Eynard-Orantin's topological recursion \cite{Eynard:2007kz,Eynard:2008we}
\begin{equation}
\begin{aligned}
 \om_{g,n+1}(z_0,J)=\sum_{\al=\pm1}\underset{z=\al}{\text{Res}}K(z_0,z)
\Biggl[&\om_{g-1,n+2}(z,\b{z},J)\\
+&\sum_{h=0}^g\sum'_{I\subset J}
\om_{h,1+|I|}(z,I)\om_{g-h,1+n-|I|}(\b{z},J\backslash I)\Biggr],
\end{aligned} 
\label{eq:top-rec}
\end{equation}
with the initial condition 
\begin{equation}
\begin{aligned}
\om_{0,2}(z_1,z_2)&=\frac{dz_1dz_2}{(z_1-z_2)^2}.
\end{aligned} 
\end{equation}
Note that $dz_i$'s are treated as bosonic variables
\begin{equation}
\begin{aligned}
 dz_idz_j=dz_jdz_i.
\end{aligned} 
\end{equation}
In \eqref{eq:top-rec}, the prime in the summation means that
$(h,I)=(0,\emptyset),(g,J)$ are excluded
and the recursion kernel $K(z_0,z)$ is given by 
\begin{equation}
\begin{aligned}
 K(z_0,z)&=-\frac{\int_{\b{z}}^z \om_{0,2}(z,z_0)}{4y(z)dx(z)}.
\end{aligned} 
\end{equation}

Following \cite{norbury2013polynomials}, we introduce
$N_{g,n}(b_1,\cdots,b_n)$ as the
expansion coefficient of $\om_{g,n}$ around $z_i=0~(i=1,\cdots,n)$
\begin{equation}
\begin{aligned}
 \om_{g,n}(z_1,\cdots,z_n)=\sum_{b_1,\cdots,b_n=1}^\infty
N_{g,n}(b_1,\cdots,b_n)\prod_{i=1}^n b_iz_i^{b_i-1}dz_i.
\end{aligned} 
\label{eq:om-N}
\end{equation}
$N_{g,n}(b_1,\cdots,b_n)$ is interpreted as a discrete analogue
of the volume of the moduli space of genus-$g$
Riemann surfaces
with $n$ boundaries \cite{norbury2013polynomials}.
For the Gaussian matrix model,
it was proved in 
\cite{norbury2008counting}
that $N_{g,n}(b_1,\cdots,b_n)$ counts the number of lattice points
on the moduli space of curves.
As shown in \cite{norbury2013polynomials},
$N_{g,n}(b_1,\cdots,b_n)$ is a symmetric polynomial in $b_i^2~(i=1,\cdots,n)$.
Although $N_{g,n}(b_1,\cdots,b_n)$
is only defined for $b_i\geq1$ 
in the original definition \eqref{eq:om-N},
there is a natural analytic continuation to $b_i=0$
since $N_{g,n}$ is a polynomial in $b_i^2$.

As discussed in \cite{Okuyama:2023kdo}, the connected correlator
of the partition function $Z(\bt)=\Tr e^{-\bt M}$ can be 
constructed by gluing the discrete volume 
and the trumpet 
\begin{equation}
\begin{aligned}
 \left\bra\prod_{i=1}^n Z(\bt_i)\right\ket_{\text{conn}}
&=\sum_{g=0}^\infty N^{2-2g-n}Z_{g,n}(\bt_1,\cdots,\bt_n),\\
Z_{g,n}(\bt_1,\cdots,\bt_n)&=\sum_{b_1,\cdots,b_n=1}^\infty
N_{g,n}(b_1,\cdots,b_n)\prod_{i=1}^n b_iZ_{\text{trumpet}}(b_i,\bt_i),
\end{aligned} 
\label{eq:Zconn}
\end{equation}
where the trumpet is given by
\begin{equation}
\begin{aligned}
 Z_{\text{trumpet}}(b,\bt)=\oint_{z=0}\frac{dz}{2\pi\ri}
z^{b-1} e^{-\bt x(z)}=e^{-\bt\cob}I_b(-2\bt\ga).
\end{aligned} 
\label{eq:trumpet}
\end{equation}
Here $I_\nu(x)$ denotes the modified Bessel function of the first kind.

As shown in \cite{Eynard:2007kz}, $\om_{g,n}$
satisfies the so-called dilation equation
\begin{equation}
\begin{aligned}
 -\sum_{\al=\pm1}\underset{z=\al}{\text{Res}}
\Bigl[\Phi(z)\om_{g,n+1}(z,z_1,\cdots,z_n)\Bigr]=(2-2g-n)
\om_{g,n}(z_1,\cdots,z_n),
\end{aligned} 
\label{eq:dilaton-om}
\end{equation}
where $\Phi(z)$ is the effective potential defined by
\begin{equation}
\begin{aligned}
 \Phi(z)=\int^z ydx.
\end{aligned} 
\end{equation}  
As a special case of \eqref{eq:dilaton-om}, 
the genus-$g$ free energy $F_g$ is obtained from $\om_{g,1}$
\begin{equation}
\begin{aligned}
 F_g=-\frac{1}{2-2g}\sum_{\al=\pm1}\underset{z=\al}{\text{Res}}
\Bigl[\Phi(z)\om_{g,1}(z)\Bigr].
\end{aligned} 
\end{equation}

\section{Cap amplitude $\psi(b)$}\label{sec:cap}
In this section, we introduce the cap amplitude $\psi(b)$ and rewrite
the dilaton equation \eqref{eq:dilaton-om} in terms of $\psi(b)$ and the 
discrete volume $N_{g,n}$.
 
We define the cap amplitude $\psi(b)$
as the expansion coefficients of the 1-form $ydx$
on the spectral curve
\begin{equation}
\begin{aligned}
 ydx=-\frac{dz}{2z}\sum_{b=0}^\infty\psi(b)(z^b+z^{-b}).
\end{aligned} 
\label{eq:ydx-psi}
\end{equation}
From the normalization of $R(x)$ in \eqref{eq:Rx},
 $R(x)$ behaves in the large $x$ limit as
\begin{equation}
\begin{aligned}
 R(x)\to\frac{1}{x}+\cO(x^{-2}),\quad (x\to\infty).
\end{aligned} 
\label{eq:R-1/x}
\end{equation}
Since $z=0$ is mapped to $x=\infty$ by the Joukowsky map
\eqref{eq:Joukowsky}, the residue of $ydx$ at $z=0$
should be normalized as
\begin{equation}
\begin{aligned}
\underset{z=0}{\text{Res}}\bigl[\,ydx\,\bigr]=-1.
\end{aligned} 
\end{equation}
This condition fixes $\psi(0)$ as
\begin{equation}
\begin{aligned}
 \psi(0)=1.
\end{aligned}
\label{eq:psi0} 
\end{equation} 
Since $ydx$ vanishes at $z=\pm1$, $\psi(b)$ should satisfy
\begin{equation}
\begin{aligned}
 \sum_{b=0}^\infty\psi(b)=\sum_{b=0}^\infty\psi(b)(-1)^b=0.
\end{aligned}
\label{eq:sum-psi}
\end{equation} 

\subsection{Dilaton equation for $N_{g,n}$}
As discussed in \cite{norbury2013polynomials}, 
the dilaton equation \eqref{eq:dilaton-om} for $\om_{g,n}$
can be recast as the dilaton equation for $N_{g,n}$
by the method of partial integration and contour deformation.
By the partial integration, \eqref{eq:dilaton-om}
is rewritten as
\begin{equation}
\begin{aligned}
 (2-2g-n)\om_{g,n}(z_1,\cdots,z_n)&=
\sum_{\al=\pm1}\underset{z=\al}{\text{Res}}
\Bigl[ydx\int^z\om_{g,n+1}(z,z_1,\cdots,z_n)\Bigr].
\end{aligned} 
\label{eq:partial1}
\end{equation}
We also perform the following contour deformation on the $z$-plane:
\begin{equation}
\begin{tikzpicture}[scale=0.7]
\draw (2,0) node [above]{$1$};
\draw (0,0) node [above]{$0$};
\draw (-2,0) node [above]{$-1$};
\draw (2,-0.85) node [below]{$C_1$};
\draw (-2,-0.85) node [below]{$C_{-1}$};
\filldraw[black] (2,0) circle (2pt);
\filldraw[black] (-2,0) circle (2pt);
\filldraw[black] (0,0) circle (2pt);
\draw[->] (2.8,0) arc (0:360:0.8);
\draw[->] (-1.2,0) arc (0:360:0.8);
\draw (4,-0.3) node [above]{$=$};
\filldraw[black] (10,0) circle (2pt);
\filldraw[black] (6,0) circle (2pt);
\filldraw[black] (8,0) circle (2pt);
\draw (10,0) node [above]{$1$};
\draw (6,0) node [above]{$-1$};
\draw (8,0) node [above]{$0$};
\draw[->] (9.5,0) arc (0:-360:1.5);
\draw[->] (10.5,0) arc (0:360:2.5);
\draw (8.5,1.3) node [below]{$-C_0$};
\draw (9.5,2.1) node [above]{$C_{\infty}$};
\end{tikzpicture}
\label{fig:contour}
\end{equation}
Here $C_{\pm1},C_0$ surround $z=\pm1,0$ counter-clockwise, 
and $-C_0$ is the orientation-reversal of $C_0$.
$C_\infty$ is a big counter-clockwise circle in the region $|z|>1$
surrounding all $z=\pm1,0$.
Using this contour deformation, we can rewrite \eqref{eq:partial1} as
the residue at $z=0,\infty$
\begin{equation}
\begin{aligned}
 (2-2g-n)\om_{g,n}(z_1,\cdots,z_n)&=\left(\frac{1}{2\pi\ri}\int_{C_1}
+\frac{1}{2\pi\ri}\int_{C_{-1}}\right)
\Biggl[ydx\int^z\om_{g,n+1}(z,z_1,\cdots,z_n)\Biggr]\\
&=\left(\frac{1}{2\pi\ri}\int_{C_\infty}
-\frac{1}{2\pi\ri}\int_{C_{0}}\right)
\Biggl[ydx\int^z\om_{g,n+1}(z,z_1,\cdots,z_n)\Biggr]\\
&=\sum_{\al=0,\infty}\underset{z=\al}{\text{Res}}
\Bigl[-ydx\int^z\om_{g,n+1}(z,z_1,\cdots,z_n)\Bigr].
\end{aligned} 
\label{eq:dilaton-res}
\end{equation}
From \eqref{eq:om-N}, the integral $\int^z\om_{g,n+1}$ becomes
\begin{equation}
\begin{aligned}
 \int^z\om_{g,n+1}(z,z_1,\cdots,z_n)&=\sum_{b,b_1,\cdots,b_n=1}^\infty 
N_{g,n+1}
(b,b_1,\cdots, b_n)z^b\prod_{i=1}^n b_iz_i^{b_i-1}dz_i\\
&\quad +\sum_{b_1,\cdots,b_n=1}^\infty \hf N_{g,n+1}
(0,b_1,\cdots, b_n)\prod_{i=1}^n b_iz_i^{b_i-1}dz_i.
\end{aligned} 
\label{eq:int-om-0}
\end{equation}
Similarly, from the relation 
\begin{equation}
\begin{aligned}
 \om_{g,n+1}(\b{z},z_1,\cdots,z_n)=-\om_{g,n+1}(z,z_1,\cdots,z_n),
\end{aligned} 
\end{equation}
the expansion of $\int^z\om_{g,n+1}$ around $z=\infty$ is given by
\begin{equation}
\begin{aligned}
 \int^z\om_{g,n+1}(z,z_1,\cdots,z_n)&=
-\sum_{b,b_1,\cdots,b_n=1}^\infty 
N_{g,n+1}
(b,b_1,\cdots, b_n)z^{-b}\prod_{i=1}^n b_iz_i^{b_i-1}dz_i\\
&\quad -\sum_{b_1,\cdots,b_n=1}^\infty \hf N_{g,n+1}
(0,b_1,\cdots, b_n)\prod_{i=1}^n b_iz_i^{b_i-1}dz_i.
\end{aligned} 
\label{eq:int-om-infty}
\end{equation}
In \eqref{eq:int-om-0} and \eqref{eq:int-om-infty}, 
we fixed the integration constant by the condition \cite{norbury2013polynomials}
\begin{equation}
\begin{aligned}
 \int_\infty^0\om_{g,n+1}(z,z_1,\cdots,z_n)=
\sum_{b_1,\cdots,b_n=1}^\infty  N_{g,n+1}
(0,b_1,\cdots, b_n)\prod_{i=1}^n b_iz_i^{b_i-1}dz_i.
\end{aligned} 
\end{equation}
Finally, plugging $ydx$ in \eqref{eq:ydx-psi} and
$\int^z\om_{g,n+1}$ in \eqref{eq:int-om-0}, \eqref{eq:int-om-infty}
into the last line of \eqref{eq:dilaton-res}
and taking the residue at $z=0,\infty$, 
we obtain the desired relation \eqref{eq:dilaton}.
The relation \eqref{eq:conj}
between the free energy $F_g$ and $N_{g,1}$ can be proved in a similar manner.

\subsection{The eigenvalue density $\rho_0(x)$ and the potential $V(x)$}
In this subsection, we will show that the genus-zero eigenvalue density 
$\rho_0(x)$ and the
potential $V(x)$ are written in terms of the cap amplitude $\psi(b)$.
Since the cut $[a,b]$ on the $x$-plane is mapped to the unit circle on the
$z$-plane, it is convenient to 
parameterize a point on the cut by
$z=e^{\ri\th}$.
Note that the branch points $z=1$ and $z=-1$ correspond to $\th=0$ and 
$\th=\pi$, respectively.
Then $x(z)$ in \eqref{eq:Joukowsky} becomes
\begin{equation}
\begin{aligned}
 x(e^{\ri\th})=2\ga\cos\th+\cob.
\end{aligned} 
\label{eq:x-th}
\end{equation}
It is well-known that the eigenvalue density $\rho_0(x)$ 
is obtained by
taking the discontinuity of the resolvent
$R(x)$ along the cut.
Since $R(x)$ and $y$ are related by \eqref{eq:y-def},
$\rho_0(x)$ is given by
\begin{equation}
\begin{aligned}
 \rho_0(x)dx&=\frac{1}{\pi}\text{Im}\Bigl(-ydx\big|_{z=e^{\ri\th}}\Bigr).
\end{aligned}
\label{eq:rho-mu1}
\end{equation}
Plugging \eqref{eq:ydx-psi} into \eqref{eq:rho-mu1}, we find
\begin{equation}
\begin{aligned}
 \rho_0(x)dx&=
\mu(\th)\frac{d\th}{2\pi},
\end{aligned}
\label{eq:rho-mu}
\end{equation}
where $\mu(\th)$ is given by\footnote{For $\mu(\th)$ to be positive
in the region $\th\in[0,\pi]$, we need to impose some condition
on $\psi(b)$. We will not discuss this condition in full generality.
For the example of models considered in section \ref{sec:example}, 
$\mu(\th)$ is indeed positive definite.}
\begin{equation}
\begin{aligned}
 \mu(\th)&=\sum_{b=0}^\infty \psi(b)2\cos(b\th).
\end{aligned} 
\label{eq:mu-psi}
\end{equation}
Thus we arrive at a very simple expression 
\eqref{eq:mu-psi} of the
eigenvalue density $\mu(\th)$ in terms of the variable $\th$.
From \eqref{eq:sum-psi}, the density $\mu(\th)$ vanishes at the branch points
$\th=0,\pi$.
Also, from \eqref{eq:psi0} the density \eqref{eq:mu-psi} is unit-normalized
\begin{equation}
\begin{aligned}
 \int_0^\pi\frac{d\th}{2\pi}\mu(\th)=1.
\end{aligned} 
\end{equation}
We can also write down a simple expression of the matrix model
potential $V(x)$. 
The potential satisfies the following saddle-point equation
\begin{equation}
\begin{aligned}
 \hf V'(x)=\mathcal{P}\int_{a}^b dx'\rho_0(x')\frac{1}{x-x'},
\end{aligned} 
\label{eq:V-rho}
\end{equation}
where $x$ is on the cut $[a,b]$ 
and $\mathcal{P}$ denotes the principal value.
From \eqref{eq:x-th} and \eqref{eq:rho-mu},  \eqref{eq:V-rho} is written as
\begin{equation}
\begin{aligned}
 \hf V'(x)=\mathcal{P}\int_{0}^\pi\frac{d\phi}{2\pi}\mu(\phi)
\frac{1}{2\ga\cos\th-2\ga\cos\phi}.
\end{aligned} 
\end{equation}
Using the formula 
\begin{equation}
\begin{aligned}
 \mathcal{P}\int_0^\pi\frac{d\phi}{2\pi}\frac{2\cos(b\phi)}{\cos\th-\cos\phi}
=-\frac{\sin(b\th)}{\sin\th}
\end{aligned} 
\end{equation}
we find
\begin{equation}
\begin{aligned}
 V'(x)=-\frac{1}{\ga}\sum_{b=1}^\infty \psi(b)
\frac{\sin(b\th)}{\sin\th}=-\frac{1}{\ga}\sum_{b=1}^\infty \psi(b)
U_{b-1}\left(\frac{x-\cob}{2\ga}\right),
\end{aligned}
\label{eq:Vprime} 
\end{equation}
where $U_n(\cos\th)=\frac{\sin(n+1)\th}{\sin\th}$ denotes the Chebyshev
polynomial of the second kind.
By integrating \eqref{eq:Vprime}, we arrive at
the desired expression of $V(x)$ in terms of $\psi(b)$
\begin{equation}
\begin{aligned}
 V(x)
=-\sum_{b=1}^\infty\frac{1}{b}\psi(b)2\cos(b\th)
=-\sum_{b=1}^\infty\frac{1}{b}\psi(b)2T_b\left(\frac{x-\cob}{2\ga}\right),
\end{aligned} 
\label{eq:V-psi}
\end{equation}
where $T_n(\cos\th)=\cos (n\th)$ denotes 
the Chebyshev
polynomial of the first kind.

Note that \eqref{eq:mu-psi} and \eqref{eq:V-psi} have been
already considered in \cite{Borot:2010tr} (see (3-13) and (3-20)
in \cite{Borot:2010tr}). Their $v_b$ and our $\psi(b)$
are related by
\begin{equation}
\begin{aligned}
 \psi(b)=bv_b,\quad (b\geq1).
\end{aligned} 
\label{eq:psi-vb}
\end{equation}
Our novel finding is that $\psi(b)$
has a geometric interpretation as the cap amplitude
with the boundary length $b$, which naturally appears in the dilaton equation
\eqref{eq:dilaton}.

The constant term $v_0$ in the potential of \cite{Borot:2010tr} 
is set to zero in our expression \eqref{eq:V-psi}.
We should stress that our $\psi(0)$ is not directly related to $v_0$ in \cite{Borot:2010tr}; $v_0$ in \cite{Borot:2010tr}
is a constant term in the potential $V(x)$ while our $\psi(0)$
is the residue of the 1-form $-ydx$ at $z=0$.  
We cannot determine the value of $\psi(0)$ from the 
potential since $\psi(0)$ does not appear in \eqref{eq:V-psi}.
As we mentioned in section \ref{sec:intro}, this is the 
reason why we use the 1-form $ydx$ to define
the cap amplitude $\psi(b)$.

As an application of our formalism, let us consider the disk partition
function $Z_{\text{disk}}(\bt)$, which corresponds to $(g,n)=(0,1)$ in 
\eqref{eq:Zconn}. $Z_{\text{disk}}(\bt)$
is written as the average of $e^{-\bt x}$ with respect to the genus-zero 
eigenvalue
density $\rho_0(x)$
\begin{equation}
\begin{aligned}
 Z_{\text{disk}}(\bt)=\int dx \rho_0(x)e^{-\bt x}=\int_0^\pi\frac{d\th}{2\pi}
\mu(\th)e^{-\bt x(e^{\ri\th})}.
\end{aligned} 
\end{equation}
Using $\mu(\th)$ in \eqref{eq:mu-psi}, we find
\begin{equation}
\begin{aligned}
 Z_{\text{disk}}(\bt)=\sum_{b=0}^\infty\psi(b)Z_{\text{trumpet}}(b,\bt),
\end{aligned} 
\label{eq:Zdisk-psi}
\end{equation}
where the trumpet is given by
\begin{equation}
\begin{aligned}
 Z_{\text{trumpet}}(b,\bt)=\int_0^\pi\frac{d\th}{2\pi}2\cos(b\th)
e^{-\bt x(e^{\ri\th})}=e^{-\bt\cob}I_b(-2\bt\ga).
\end{aligned} 
\end{equation}
This agrees with \eqref{eq:trumpet} as expected.
The expansion of $Z_\text{disk}(\bt)$ in \eqref{eq:Zdisk-psi}
is schematically depicted as
\begin{equation}
\begin{aligned}
\begin{tikzpicture}[scale=0.7]
\draw[thick,red] (1,0.5) arc [start angle=90,end angle=270, x radius=0.25, y radius=0.5];
\draw[thick,dashed,red] (1,0.5) arc [start angle=90,end angle=-90, x radius=0.25, y radius=0.5];
\draw (1,-0.5) node [below]{$b$};
\draw (1,0.5) .. controls (-0.5,0.3) and (-0.5,-0.3) .. (1,-0.5);
\draw (-0.4,0) node [left]{$\displaystyle Z_\text{disk}(\bt)~=~\sum_{b=0}^\infty$};
\draw (1,0.5) to [out=10,in=210] (4,1.5);
\draw (1,-0.5) to [out=-10,in=-210] (4,-1.5);
\draw[thick,blue] (4.4,0) arc [start angle=360,end angle=0, x radius=0.4, y radius=1.5];
\draw (4,-1.5) node [below]{$\bt$};
\end{tikzpicture}
\end{aligned}
\label{fig:Zdisk}
\end{equation}
This expansion \eqref{eq:Zdisk-psi} of $Z_\text{disk}(\bt)$ in terms of the 
trumpet has been studied in
\cite{Okuyama:2024eyf} in the case of the ETH matrix model for DSSYK.
Our result \eqref{eq:Zdisk-psi} is a generalization of 
\cite{Okuyama:2024eyf} to the matrix model with an arbitrary potential.

\subsection{Moments $M_k$ and $J_k$}
As discussed in \cite{Ambjorn:1992gw},
the free energy $F_g$ is given by some combination of the moments
$M_k$ and $J_k$\footnote{
We call $M_k,J_k$ as ``moments''
following the terminology in \cite{Ambjorn:1992gw}. We should stress that 
they are not equal 
to the moment $\bra\Tr M^k\ket$ of the matrix $M$.
Rather, they are defined by the coefficients in \eqref{eq:Q-exp}.}, which appear as the expansion coefficients
of $Q(x)$ in \eqref{eq:y-Q} around the branch points $x=a,b$
\begin{equation}
\begin{aligned}
 Q(x)=\sum_{k=1}^\infty M_k(x-a)^{k-1}=\sum_{k=1}^\infty J_k(x-b)^{k-1}.
\end{aligned} 
\label{eq:Q-exp}
\end{equation}
In this subsection, we will show that 
$M_k$ and $J_k$ are written in terms of the cap amplitude $\psi(b)$.

Using \eqref{eq:xa-xb}, \eqref{eq:root-z}, and
\begin{equation}
\begin{aligned}
 dx(z)=\ga(z-z^{-1})\frac{dz}{z},
\end{aligned} 
\end{equation}
the 1-form $ydx$ is expanded as
\begin{equation}
\begin{aligned}
 ydx&=\hf Q(x)\rt{(x-a)(x-b)}dx\\
&=\frac{dz}{2z}\sum_{k=1}^\infty M_k\ga^{k+1}(z+z^{-1}-2)^{k-1}(z-z^{-1})^2\\
&=\frac{dz}{2z}\sum_{k=1}^\infty J_k\ga^{k+1}(z+z^{-1}+2)^{k-1}(z-z^{-1})^2.
\end{aligned} 
\label{eq:ydx-MJ}
\end{equation}
Let us first consider the expansion around $x=a$.
Setting $z=e^{\ri\th}$, the second line of
\eqref{eq:ydx-MJ} becomes
\begin{equation}
\begin{aligned}
  ydx\big|_{z=e^{\ri\th}}
=\frac{\ri d\th}{2}\sum_{k=1}^\infty M_k\ga^{k+1}(2\cos\th-2)^{k-1}(-4\sin^2\th).
\end{aligned} 
\label{eq:ydx-M}
\end{equation}
On the other hand, $ydx$ 
is also written in terms of the cap amplitude $\psi(b)$.
From \eqref{eq:ydx-psi} we find
\begin{equation}
\begin{aligned}
  ydx\big|_{z=e^{\ri\th}}=-\frac{\ri d\th}{2}\sum_{b=0}^\infty\psi(b)2\cos(b\th).
\end{aligned} 
\end{equation}
Using \eqref{eq:sum-psi}, this is further rewritten as
\begin{equation}
\begin{aligned}
  ydx\big|_{z=e^{\ri\th}}&=-\frac{\ri d\th}{2}\sum_{b=0}^\infty\psi(b)(2\cos(b\th)-2)\\
&=-\frac{\ri d\th}{2}\sum_{b=0}^\infty\psi(b)\Bigl(-4\sin^2\frac{b}{2}\th\Bigr).
\end{aligned}
\label{eq:ydx-sin} 
\end{equation}
From \eqref{eq:ydx-M} and \eqref{eq:ydx-sin}
 we arrive at the following relation
\begin{equation}
\begin{aligned}
\sum_{k=1}^\infty  M_k\ga^{k+1}(2\cos\th-2)^{k-1}
&=-\sum_{b=0}^\infty \psi(b)\frac{\sin^2\frac{b}{2}\th}{\sin^2\th}.
\end{aligned} 
\end{equation}
Similarly, by equating the two expression of
$ydx$ for $z=-e^{\ri\th}$ around $x=b$, we find
\begin{equation}
\begin{aligned}
 J_k\ga^{k+1}(-2\cos\th+2)^{k-1}
&=-\sum_{b=0}^\infty \psi(b)(-1)^b\frac{\sin^2\frac{b}{2}\th}{\sin^2\th}.
\end{aligned} 
\end{equation}
Finally, we arrive at the desired relation between the moments $M_k,J_k$ and
the cap amplitude $\psi(b)$
\begin{equation}
\begin{aligned}
 M_k\ga^{k+1}&=-\sum_{b=0}^\infty\psi(b)c_k(b),\\
J_k(-\ga)^{k+1}&=-\sum_{b=0}^\infty\psi(b)(-1)^bc_k(b),
\end{aligned} 
\label{eq:MJ-psi}
\end{equation}
where $c_k(b)$ is the expansion coefficient of 
$\frac{\sin^2(b\th/2)}{\sin^2\th}$
\begin{equation}
\begin{aligned}
 \frac{\sin^2\frac{b}{2}\th}{\sin^2\th}=\sum_{k=1}^\infty
c_k(b)(2\cos\th-2)^{k-1}.
\end{aligned} 
\label{eq:cb-def}
\end{equation}
From this definition, 
we can easily find the first few terms
of $c_k(b)$
\begin{equation}
\begin{aligned}
 c_1(b)&=\frac{b^2}{4},\\
c_2(b)&=\frac{b^2(b^2-4)}{2\cdot 4!},\\
c_3(b)&=\frac{b^2(b^2-4)(b^2-17/2)}{2\cdot 6!},\\
c_4(b)&=\frac{b^2(b^2-4)(b^2-8)(b^2-16)}{2\cdot 8!}.
\end{aligned} 
\end{equation}
Our result \eqref{eq:MJ-psi} shows that
the moments $M_k,J_k$ are completely 
determined by the cap amplitude $\psi(b)$, since $c_k(b)$ is 
obtained from the known function $\frac{\sin^2(b\th/2)}{\sin^2\th}$.

\subsection{Discrete volume $N_{g,n}$}
From the topological recursion
\eqref{eq:top-rec}, $\om_{g,n}$ is determined
by the coefficients of the 1-form $ydx$. Using the expansion
 of $ydx$ in \eqref{eq:ydx-MJ}, 
the resulting $\om_{g,n}$ is expressed in terms of the moments $M_k,J_k$.
For instance, $\om_{0,3}$ is given by \cite{eynard2016counting}
\begin{equation}
\begin{aligned}
 \om_{0,3}(z_1,z_2,z_3)=
\frac{1}{2M_1\ga^2}\prod_{i=1}^3\frac{dz_i}{(1-z_i)^2}
-\frac{1}{2J_1\ga^2}\prod_{i=1}^3\frac{dz_i}{(1+z_i)^2}.
\end{aligned} 
\label{eq:om03}
\end{equation}
From this result,
we can easily read off the discrete volume $N_{0,3}$
(see \eqref{eq:om-N} for the definition of $N_{g,n}$)
\begin{equation}
\begin{aligned}
 N_{0,3}(b_1,b_2,b_3)=\frac{1}{2M_1\ga^2}+\frac{(-1)^{b_1+b_2+b_3}}{2J_1\ga^2}.
\end{aligned} 
\end{equation}
In a similar manner, we find the explicit form of $N_{1,1}$ and $N_{1,2}$
\begin{equation}
\begin{aligned}
 N_{1,1}(b)&=\frac{1}{2M_1\ga^2}\left(\frac{b^2-4}{48}-\frac{M_2\ga}{8M_1}\right)
+\frac{(-1)^b}{2J_1\ga^2}\left(\frac{b^2-4}{48}
+\frac{J_2\ga}{8J_1}\right),
\end{aligned} 
\label{eq:N11}
\end{equation}
\begin{equation}
\begin{aligned}
 N_{1,2}(b_1,b_2)&=
\frac{1}{(M_1\ga^2)^2}\Biggl[
\frac{(b_1^2+b_2^2-2)(b_1^2+b_2^2-10)}{768}+\frac{3(M_1-J_1)}{512J_1}
+\frac{3\ga^2M_2^2}{16M_1^2}-\frac{\ga J_2M_1}{128J_1^2}\\
&\quad \quad-\frac{(b_1^2+b_2^2-3)\ga M_2+5\ga^2M_3}{32M_1}\Biggr]
+\frac{(-1)^{b_1+b_2}}{(J_1\ga^2)^2}\Bigl[M_k\leftrightarrow (-1)^{k+1}J_k\Bigr]\\
&\quad +\frac{1}{128}
\left(\frac{1}{M_1\ga^2}+\frac{(-1)^{b_1}}{J_1\ga^2}\right)\left(
\frac{1}{M_1\ga^2}+\frac{(-1)^{b_2}}{J_1\ga^2}\right).
\end{aligned} 
\end{equation}
Using \eqref{eq:sum-psi} and \eqref{eq:MJ-psi},
one can check that the above
result of $N_{1,1}$ and $N_{1,2}$ indeed satisfy the dilaton equation
\eqref{eq:dilaton}
\begin{equation}
\begin{aligned}
 \sum_{b=0}^\infty \psi(b)N_{1,2}(b,b_1)=-N_{1,1}(b_1).
\end{aligned} 
\end{equation}

\subsection{Free energy $F_g$}
In this subsection, we compute the free energy $F_g~(g=0,1,2)$ for 
the general potential $V(x)$.
The relation \eqref{eq:conj} between $F_g$ and $N_{g,1}$ is only
valid for $g\geq2$,
and hence we discuss the cases of $g=0,1$ separately.
 
\subsubsection{Genus-zero}
As discussed in \cite{Borot:2010tr},
the genus-zero free energy $F_0$ can be expressed in terms of the 
cap amplitude $\psi(b)$.
To this end, it is convenient to
change the integration variable $M$ as
\begin{equation}
\begin{aligned}
 M\to \ga M+\cob\id,
\end{aligned} 
\end{equation}
where $\id$ is the $N\times N$ identity matrix.
Then the $\ga,\cob$-dependence of 
the potential in \eqref{eq:V-psi} drops out
\begin{equation}
\begin{aligned}
 V_0(M)\equiv V(\ga M+\cob\id)=
-\sum_{b=1}^\infty\frac{1}{b}\psi(b)2T_b(M/2).
\end{aligned} 
\label{eq:V0-psi}
\end{equation}
Under this change of variable, the partition function 
becomes
\begin{equation}
\begin{aligned}
 \cZ=\int d(\ga M+\cob\id)
e^{-N\Tr V(\ga M+\cob\id)}=\ga^{N^2}\int dM e^{-N\Tr V_0(M)}.
\end{aligned} 
\end{equation}
Since $\ga^{N^2}=e^{N^2\log\ga}$, 
 $F_0$ is shifted by a constant $\log\ga$
in this change of variable.
The non-trivial part of $F_0$ comes
from the potential and the Vandermonde determinant
\begin{equation}
\begin{aligned}
 F_0&=-\int dx\rho_0(x)V_0(x)+\mathcal{P}\int dx \int dx'\rho_0(x)\rho_0(x')\log|x-x'|+\log\ga\\
&=-\hf\int dx\rho_0(x)V_0(x)+\log\ga,
\end{aligned} 
\label{eq:F0-int}
\end{equation}
where the second term of the first line is given by the principal value, and
in the second equality
we used the saddle-point equation \eqref{eq:V-rho}
with $V(x)$ replaced by $V_0(x)$.
Using \eqref{eq:mu-psi} and \eqref{eq:V0-psi},
the integral in \eqref{eq:F0-int} is evaluated as
\begin{equation}
\begin{aligned}
 -\hf\int dx\rho_0(x)V_0(x)=-\hf\int_0^\pi\frac{d\th}{2\pi}\mu(\th)
\left(-\sum_{b=1}^\infty\frac{1}{b}\psi(b)2\cos(b\th)\right)=\sum_{b=1}^\infty\frac{1}{2b}\psi(b)^2.
\end{aligned} 
\end{equation}
Finally we find \footnote{Our result \eqref{eq:F0-psi} is similar to 
the expression of $F_0$ proposed in \cite{Blommaert:2025rgw}, but the
factor of $1/2$
is missing in \cite{Blommaert:2025rgw}.}
\begin{equation}
\begin{aligned}
 F_0=\sum_{b=1}^\infty \frac{1}{2b}\psi(b)^2+\log\ga.
\end{aligned} 
\label{eq:F0-psi}
\end{equation}
Using the dictionary \eqref{eq:psi-vb},
this agrees with the result of \cite{Borot:2010tr}.
\footnote{See (3-28) in \cite{Borot:2010tr}. 
Our convention corresponds to $t=1,v_0=0$ in \cite{Borot:2010tr}.}

\subsubsection{Genus-one}
As shown in \cite{Ambjorn:1992gw}, the genus-one free energy is given by
the moments $M_1,J_1$
\begin{equation}
\begin{aligned}
 F_1=-\frac{1}{24}\log(M_1\ga^2)-\frac{1}{24}\log(J_1\ga^2).
\end{aligned} 
\label{eq:F1}
\end{equation}
Although \eqref{eq:conj} is not valid for $g=1$, it is interesting to compute
the summation $\sum_{b}\psi(b)N_{1,1}(b)$ 
appearing in \eqref{eq:conj}. 
From the structure of $N_{1,1}(b)$ in \eqref{eq:N11}, 
it is natural to decompose $N_{1,1}(b)$ into 
the contributions of $z=1$ and $z=-1$
\begin{equation}
\begin{aligned}
 N_{1,1}(b)=N_{1,1}^{+}(b)+N_{1,1}^{-}(b),
\end{aligned} 
\end{equation}
where $N_{1,1}^{\pm}(b)$ are given by
\begin{equation}
\begin{aligned}
 N_{1,1}^{+}(b)&=\frac{1}{2M_1\ga^2}\left(\frac{b^2-4}{48}-\frac{M_2\ga}{8M_1}\right),\\
N_{1,1}^{-}(b)&=\frac{(-1)^b}{2J_1\ga^2}\left(\frac{b^2-4}{48}
+\frac{J_2\ga}{8J_1}\right).
\end{aligned} 
\end{equation}
Using \eqref{eq:sum-psi} and the relation \eqref{eq:MJ-psi} for $k=1$
\begin{equation}
\begin{aligned}
 M_1\ga^2=-\sum_{b=0}^\infty\psi(b)\frac{b^2}{4},\quad\quad
J_1\ga^2=-\sum_{b=0}^\infty\psi(b)(-1)^b\frac{b^2}{4},
\end{aligned} 
\end{equation}
we find
\begin{equation}
\begin{aligned}
 \sum_{b=0}^\infty\psi(b)N_{1,1}^{+}(b)=
\sum_{b=0}^\infty\psi(b)N_{1,1}^{-}(b)=-\frac{1}{24}.
\end{aligned} 
\end{equation}

We can 
obtain $F_1$ in \eqref{eq:F1} by the following heuristic argument.
The free energy has the scaling property \cite{Eynard:2008we}
\begin{equation}
\begin{aligned}
 y\to\la y~~~\Rightarrow~~~F_g\to\la^{2-2g}F_g.
\end{aligned} 
\end{equation}
Since $M_1\ga^2$ and $J_1\ga^2$ appear as the overall normalization
of $ydx$  in \eqref{eq:ydx-MJ},
it is natural to expect that \eqref{eq:conj} for $g=1$
should be understood as the following $g\to1$ limit
\begin{equation}
\begin{aligned}
 F_1&=\lim_{g\to1}\frac{1}{2-2g}\left[(M_1\ga^2)^{2-2g}
\sum_{b=0}^\infty\psi(b)N_{1,1}^{+}(b)
+(J_1\ga^2)^{2-2g}
\sum_{b=0}^\infty\psi(b)N_{1,1}^{-}(b)\right]\\
&=-\frac{1}{12}\frac{1}{2-2g}-\frac{1}{24}\log(M_1\ga^2)
-\frac{1}{24}\log(J_1\ga^2)+\cO\bigl((2-2g)\bigr).
\end{aligned} 
\label{eq:heuristic}
\end{equation}
The finite terms reproduce \eqref{eq:F1}.
The first divergent term in 
\eqref{eq:heuristic} is just an artifact of this heuristic derivation.
We should stress that the correct $F_1$ is finite
as shown in \cite{Ambjorn:1992gw}. 

\subsubsection{Genus-two}
Next, let us consider the genus-two free energy $F_2$.
For $g=2$, we can safely use the dilaton equation \eqref{eq:conj} to compute 
$F_2$ from $N_{2,1}(b)$.
Using the topological recursion, we find
\begin{equation}
\begin{aligned}
 N_{2,1}(b)=\frac{1}{2(M_1\ga^2)^3}\left(a_0+
\sum_{k=1}^4a_kc_k(b)\right)
+\frac{(-1)^b}{2(J_1\ga^2)^3}\left(\til{a}_0+
\sum_{k=1}^4\til{a}_kc_k(b)\right),
\end{aligned} 
\label{eq:N21-result}
\end{equation}
where $c_k(b)$ is defined in \eqref{eq:cb-def} and
$a_k~(k=1,\cdots,4)$ are given by 
\begin{equation}
\begin{aligned}
 a_4&=\frac{35}{768},\\
a_3&=-\frac{29 \gamma  M_2}{256 M_1}-\frac{43}{1536},\\
a_2&=
-\frac{\gamma  J_2 M_1}{512 J_1^2}+\frac{3 M_1}{2048 J_1}+\frac{63 \gamma ^2 M_2^2}{320
   M_1^2}-\frac{29 \gamma ^2 M_3}{256 M_1}+\frac{11 \gamma  M_2}{160 M_1}+\frac{181}{15360},\\
a_1&=\frac{\gamma ^2 J_2 M_2}{256 J_1^2}+\frac{3 \gamma  J_2 M_1}{2048 J_1^2}-\frac{3 \gamma 
   M_2}{1024 J_1}-\frac{5 M_1}{2048 J_1}-\frac{21 \gamma ^3 M_2^3}{64 M_1^3}+\frac{29 \gamma ^3
   M_2 M_3}{64 M_1^2}\\
&\quad -\frac{35 \gamma ^3 M_4}{256 M_1}-\frac{11 \gamma ^2 M_2^2}{80
   M_1^2}+\frac{43 \gamma ^2 M_3}{512 M_1}-\frac{181 \gamma  M_2}{5120 M_1}-\frac{181}{30720}.
\end{aligned} 
\label{eq:ak}
\end{equation}
$a_0$ is some complicated combination of the moments.
$\til{a}_k$'s are obtained from $a_k$ by
\begin{equation}
\begin{aligned}
 \til{a}_k=a_k\Big|_{M_i\leftrightarrow (-1)^{i+1}J_i}.
\end{aligned} 
\label{eq:til-ak}
\end{equation}

From the dilaton equation \eqref{eq:conj}, $F_2$ is given by
\begin{equation}
\begin{aligned}
 F_2=-\hf\sum_{b=0}^\infty\psi(b)N_{2,1}(b).
\end{aligned} 
\label{eq:F2-sum}
\end{equation}
Plugging \eqref{eq:N21-result} into \eqref{eq:F2-sum} 
and using \eqref{eq:MJ-psi}, $F_2$  becomes
\begin{equation}
\begin{aligned}
 F_2
&=\frac{1}{4(M_1\ga^3)^3}\sum_{k=1}^4 a_kM_k\ga^{k+1}
+\frac{1}{4(J_1\ga^3)^3}\sum_{k=1}^4 \til{a}_kJ_k(-\ga)^{k+1}.
\end{aligned} 
\label{eq:F2-result}
\end{equation}
Note that $a_0$ and $\til{a}_0$ do not contribute to $F_2$
because of the relation \eqref{eq:sum-psi}.
Using the explicit form of $a_k,\til{a}_k$ in \eqref{eq:ak} with 
\eqref{eq:til-ak},
one can check that \eqref{eq:F2-result} 
agrees with the result of 
\cite{Ambjorn:1992gw}.\footnote{
$d$ in \cite{Ambjorn:1992gw} and our $\ga$ are related by $\ga=d/4$ 
(see \eqref{eq:ga-cob}).
Note that $F_2$ in the original version of \cite{Ambjorn:1992gw} contains an error.
See the erratum of \cite{Ambjorn:1992gw} for the correct $F_2$.}

\section{Examples}\label{sec:example}
In this section, we consider 
the Gaussian matrix model and the ETH matrix for DSSYK
as examples of our formalism.
For those cases, the potential $V(x)$ is an even function
\begin{equation}
\begin{aligned}
 V(-x)=V(x).
\end{aligned} 
\end{equation}
In this case, some of the formulas in the previous section 
simplify. 
The end-points of the cut $[a,b]$ satisfy $b=-a$,
since the eigenvalues 
are distributed symmetrically around the origin $x=0$
for the even potential case.
Then $\cob$ in \eqref{eq:ga-cob} vanishes.
When $\cob=0$, the Joukowsky map
\eqref{eq:Joukowsky} becomes
\begin{equation}
\begin{aligned}
 x(z)=\ga(z+z^{-1}),
\end{aligned} 
\end{equation} 
and the potential in \eqref{eq:V-psi} becomes
\begin{equation}
\begin{aligned}
 V(x)=-\sum_{b=1}^{\infty}\frac{1}{b}\psi(b)2T_b(x/2\ga).
\end{aligned} 
\end{equation}
Since $T_b(-x)=(-1)^b T_b(x)$,  $\psi(b)$
should vanish for odd $b$ in the even potential case
\begin{equation}
\begin{aligned}
 \psi(b)=0,\quad (b=\text{odd}).
\end{aligned} 
\end{equation}
In this case, the moments $M_k$ and $J_k$ are not independent. 
From \eqref{eq:MJ-psi}, they are related by
\begin{equation}
\begin{aligned}
 J_k=(-1)^{k+1}M_k.
\end{aligned} 
\label{eq:J-M}
\end{equation}
Using this relation, the expression 
of the discrete volume $N_{g,n}$ obtained in the previous section
simplifies to
\begin{equation}
\begin{aligned}
N_{0,3}(b_1,b_2,b_3)&=\frac{1}{M_1\ga^2}P_{b_1+b_2+b_3},\\
 N_{1,1}(b)&=\frac{1}{M_1\ga^2}\left(\frac{b^2-4}{48}
-\frac{M_2\ga}{8M_1}\right)P_b,\\
N_{1,2}(b_1,b_2)&=\frac{1}{(M_1\ga^2)^2}
\Biggl[
\frac{(b_1^2+b_2^2-2)(b_1^2+b_2^2-10)}{384}
-\frac{(4b_1^2+4b_2^2-13)\ga M_2}{64M_1}\\
&\hskip20mm+
\frac{3\ga^2M_2^2}{8M_1^2}-\frac{5\ga^2M_3}{16M_1}\Biggr]P_{b_1+b_2}
+\frac{P_{b_1}P_{b_2}}{32(M_1\ga^2)^2},
\end{aligned} 
\label{eq:Ngn-even}
\end{equation}
where $P_b$ is the projection to even $b$
\begin{equation}
\begin{aligned}
 P_b=\frac{1+(-1)^b}{2}.
\end{aligned} 
\end{equation}
Note that 
$N_{g,n}$ in \eqref{eq:Ngn-even}
is proportional to $(M_1\ga^2)^{2-2g-n}$, which
follows from the scaling property \cite{Eynard:2008we}
\begin{equation}
\begin{aligned}
 y\to\la y~~~\Rightarrow~~~\om_{g,n}\to\la^{2-2g-n}\om_{g,n}.
\end{aligned} 
\end{equation}
 
\subsection{Gaussian matrix model}\label{sec:gaussian}
As the first example,
let us consider the Gaussian matrix model
\begin{equation}
\begin{aligned}
 \cZ_G=\int dM e^{-\frac{N}{2}\Tr M^2}.
\end{aligned} 
\label{eq:ZGauss}
\end{equation}
In this model,
the parameters $\ga,\cob$ in the Joukowsky map
\eqref{eq:Joukowsky} and $Q(x)$ in \eqref{eq:y-Q} are given by
\begin{equation}
\begin{aligned}
 \ga=1,\quad \cob=0,\quad Q(x)=1.
\end{aligned} 
\end{equation}
Then the spectral curve becomes
\begin{equation}
\begin{aligned}
 x(z)=z+z^{-1},\quad y(z)=\hf (z-z^{-1}),
\end{aligned} 
\end{equation}
and the 1-form $ydx$ is given by
\begin{equation}
\begin{aligned}
 ydx=-\frac{dz}{2z}(2-z^2-z^{-2}).
\end{aligned} 
\label{eq:1-Gauss}
\end{equation}
We can read off the cap amplitude $\psi(b)$ for the Gaussian matrix model
from the definition of $\psi(b)$ in \eqref{eq:ydx-psi}, 
\begin{equation}
\psi(b)=\left\{
\begin{aligned}
 1,&\qquad &&(b=0),\\
-1,&\qquad &&(b=2),\\
0,&\qquad &&(\text{otherwise}).
\end{aligned} \right.
\label{eq:psi-Gauss}
\end{equation}

Using the dilaton equation \eqref{eq:conj}, we can relate 
$N_{g,1}(b)$ and the free energy $F_g$.
As shown in \cite{norbury2008counting}, the values of 
$N_{g,1}(b)$ at $b=0,2$ are given by
\begin{equation}
\begin{aligned}
 N_{g,1}(0)=\zeta(1-2g)=-\frac{B_{2g}}{2g},\quad N_{g,1}(2)=0,
\end{aligned} 
\end{equation}
where $B_{2g}$ denotes the Bernoulli number.
From \eqref{eq:conj} the genus-$g$ free energy becomes
\begin{equation}
\begin{aligned}
 F_g=\frac{N_{g,1}(0)-N_{g,1}(2)}{2-2g}=\frac{B_{2g}}{2g(2g-2)},\quad(g\geq2).
\end{aligned} 
\label{eq:Fg-gauss}
\end{equation}
As discussed in \cite{Ooguri:2002gx},
this agrees with the free energy coming from the volume of the unitary group
$U(N)$.
$\cZ_G$ in \eqref{eq:ZGauss} is given by
\begin{equation}
\begin{aligned}
 \cZ_G=\frac{1}{\text{vol}\bigl(U(N)\bigr)}
\Biggl(\frac{2\pi}{N}\Biggr)^{\frac{N^2}{2}}.
\end{aligned} 
\label{eq:ZGauss-vol}
\end{equation}
The last factor comes from the naive Gaussian integral in \eqref{eq:ZGauss},
but we have to divide by the volume of the unitary group $U(N)$
\begin{equation}
\begin{aligned}
 \text{vol}\bigl(U(N)\bigr)=\frac{(2\pi)^{\hf N(N+1)}}{G_2(N+1)},
\end{aligned} 
\label{eq:volUN}
\end{equation}
since $\cZ_G$ has a gauge symmetry $M\to UMU^{-1}~(U\in U(N))$.
Here
$G_2(N+1)$ denotes the Barnes $G$-function. Plugging \eqref{eq:volUN} into
\eqref{eq:ZGauss-vol}, we find
\begin{equation}
\begin{aligned}
 \cZ_G=N^{-\hf N^2}(2\pi)^{-\hf N}G_2(N+1).
\end{aligned} 
\label{eq:ZGauss-Barnes}
\end{equation}
Using the asymptotic expansion of the Barnes $G$-function, 
the free energy $\log \cZ_G$ is expanded 
as
\begin{equation}
\begin{aligned}
 \log \cZ_G
=-\frac{3}{4}N^2-\frac{1}{12}\log N+\zeta'(-1)
+\sum_{g=2}^\infty \frac{B_{2g}}{2g(2g-2)} N^{2-2g}.
\end{aligned}
\label{eq:logZG} 
\end{equation}
We can see that the coefficient of $N^{2-2g}$ agrees with \eqref{eq:Fg-gauss} for 
$g\geq2$, as expected.

The genus-one free energy can be obtained by the following heuristic derivation,
in a similar manner as \eqref{eq:heuristic}
\begin{equation}
\begin{aligned}
 F_1&=\lim_{g\to1}\frac{N^{2-2g}}{2-2g}\sum_{b=0}^\infty \psi(b)N_{g,1}(b)\\
&=\lim_{g\to1}\frac{N^{2-2g}}{2-2g}\zeta(1-2g)\\
&=\frac{\zeta(-1)}{2-2g}+\zeta(-1)\log N+\zeta'(-1)+\cO\bigl((2-2g)\bigr).
\end{aligned} 
\label{eq:F1-lim}
\end{equation}
Using $\zeta(-1)=-\frac{1}{12}$,
the finite terms of \eqref{eq:F1-lim}
reproduce the genus-one part of \eqref{eq:logZG}.

The genus-zero free energy of the Gaussian matrix model is a bit tricky.
In order to make contact with the general formula \eqref{eq:F0-psi},
we have to use the potential
\begin{equation}
\begin{aligned}
 V(M)=T_2(M/2)=\frac{M^2}{2}-\id.
\end{aligned} 
\end{equation}
The partition function for this potential differs from
$\cZ_G$ in \eqref{eq:ZGauss} by an overall normalization 
\begin{equation}
\begin{aligned}
 \cZ_G'=\int dM e^{-N\Tr T_2(M/2)}=e^{N^2}\cZ_G.
\end{aligned} 
\end{equation}
Thus the free energy $\log \cZ_G'$ is shifted from \eqref{eq:logZG}
by $N^2$
\begin{equation}
\begin{aligned}
 \log \cZ_G'=N^2+\log \cZ_G=\qu N^2+\cO(N^0).
\end{aligned} 
\end{equation}
As expected, the genus zero part of $\log \cZ_G'$ 
agrees with the general formula \eqref{eq:F0-psi}
with the cap amplitude in \eqref{eq:psi-Gauss} and $\ga=1$
\begin{equation}
\begin{aligned}
 F_0=\sum_{b=1}^\infty\frac{1}{2b}\psi(b)^2=\qu \psi(2)^2=\qu.
\end{aligned} 
\end{equation}

\subsection{ETH matrix model for DSSYK}\label{sec:DSSYK}
As a next example, let us consider the ETH matrix model for DSSYK
\cite{Jafferis:2022wez}.
The SYK model is a system of $N_f$ Majorana fermions 
$\psi_i~(i=1,\cdots,N_f)$ with the $p$-body interaction \cite{Sachdev1993,Kitaev1,Kitaev2}
\begin{equation}
\begin{aligned}
 H=\ri^{p/2}\sum_{i_1<\cdots<i_p}J_{i_1\cdots i_p}\psi_{i_1}\cdots\psi_{i_p}, 
\end{aligned} 
\end{equation}
where $J_{i_1\cdots i_p}$ is a Gaussian 
random coupling with zero mean and the variance
\begin{equation}
\begin{aligned}
 \bra J_{i_1\cdots i_p}^2\ket=\mathcal{J}^2\binom{N_f}{p}^{-1}.
\end{aligned} 
\end{equation} 
DSSYK is defined by the 
following scaling limit \cite{Berkooz:2018jqr}
\begin{equation}
\begin{aligned}
 N_f,p\to\infty\quad\text{with}\quad\la=\frac{2p^2}{N_f}: \text{fixed}.
\end{aligned} 
\end{equation} 
In this limit, computation of the random average of the moment $\bra\tr H^k\ket$
boils down to the counting problem of the
chord diagrams, which can be solved by the $q$-deformed oscillator
with $q=e^{-\la}$.
As shown in \cite{Berkooz:2018jqr},
the disk partition function of DSSYK is written as 
\begin{equation}
\begin{aligned}
 Z_\text{disk}(\bt)=\int_0^\pi\frac{d\th}{2\pi}\mu(\th)e^{-\bt E(\th)},
\end{aligned} 
\end{equation}
where  $E(\th)$ and $\mu(\th)$ are given by
\begin{align}
\quad E(\th)&=2\ga \cos\th,\quad \ga=\frac{\mathcal{J}}{\rt{1-q}},\\
\mu(\th)&=(q;q)_\infty(e^{2\ri\th};q)_\infty(e^{-2\ri\th};q)_\infty.
\label{eq:mu-DSSYK} 
\end{align}
Here $(a;q)_\infty$ denotes the $q$-Pochhammer symbol 
\begin{equation}
\begin{aligned}
 (a;q)_\infty=\prod_{n=0}^\infty(1-aq^n).
\end{aligned} 
\end{equation}
The ETH matrix model for DSSYK is a hermitian one-matrix model
\eqref{eq:hermitian-mat} whose potential
$V(M)$ is constructed in such a way that 
the density \eqref{eq:mu-DSSYK} of DSSYK is reproduced in the large $N$ limit.
Here the size $N$ of the matrix is identified as the dimension of the 
Hilbert space of $N_f$ Majorana fermions
\begin{equation}
\begin{aligned}
 N=2^{N_f/2}.
\end{aligned} 
\end{equation}
From the expansion of $\mu(\th)$ in \eqref{eq:mu-DSSYK} \cite{Okuyama:2024eyf}
\begin{equation}
\begin{aligned}
 \mu(\th)=2+\sum_{r=1}^\infty (-1)^rq^{\hf r^2}(q^{\hf r}+q^{-\hf r})2\cos(2r\th),
\end{aligned}
\label{eq:mu-ETH-exp} 
\end{equation}
we can read off the cap amplitude of the ETH matrix model
by comparing \eqref{eq:mu-ETH-exp} with \eqref{eq:mu-psi}
\begin{equation}
\psi(b)=\left\{
\begin{aligned}
 &1,\quad &&(b=0),\\
&(-1)^rq^{\hf r^2}(q^{\hf r}+q^{-\hf r}),\quad &&(b=2r>0),\\
&0,\quad &&(b=\text{odd}).
\end{aligned} \right.
\label{eq:psi-ETH}
\end{equation}
Since $\psi(b)$ vanishes for odd $b$, 
the potential $V(M)$ of the ETH matrix model is an even function of $M$.
When $q=0$, the cap amplitude in \eqref{eq:psi-ETH} reduces to that
of the Gaussian matrix model in \eqref{eq:psi-Gauss}.
This is consistent with the fact that 
the ETH matrix model reduces to the Gaussian matrix model when $q=0$
\cite{Jafferis:2022wez}.

We can compute $N_{g,n}$ and $F_g$ of the ETH matrix model
using the data of $\psi(b)$ in \eqref{eq:psi-ETH}.
The moment $M_k$ is obtained by the formula \eqref{eq:MJ-psi}
and $J_k$ is related to $M_k$ by \eqref{eq:J-M}.
The fist moment $M_1$ is given by
\begin{equation}
\begin{aligned}
  M_1\ga^2=-\sum_{b=1}^\infty \psi(b)\frac{b^2}{4}=
\sum_{r=1}^\infty (-1)^{r-1}r^2\bigl(q^{\hf r(r+1)}+q^{\hf r(r-1)}\bigr).
\end{aligned} 
\end{equation}
By shifting $r\to r+1$ in the last term and
using the formula
\begin{equation}
\begin{aligned}
 \sum_{r=0}^\infty (-1)^r (2r+1)q^{\hf r(r+1)}=(q;q)_\infty^3,
\end{aligned} 
\end{equation}
we find
\begin{equation}
\begin{aligned}
 M_1\ga^2=(q;q)_\infty^3.
\end{aligned} 
\label{eq:M1-ETH}
\end{equation}
As for the higher moments,
it turns out that they are given by some combination of the
$q$-deformed zeta-function
\begin{equation}
\begin{aligned}
 \zeta_q(s)=\sum_{n=1}^\infty \bigl(q^{-\hf n}-q^{\hf n}\bigr)^{-s}.
\end{aligned} 
\end{equation}
After some algebra, we find the first few moments 
\begin{equation}
\begin{aligned}
\frac{M_2\ga}{M_1}&=-4\zeta_q(2),\\
\frac{M_3\ga^2}{M_1}&=-\zeta_q(2)+8\bigl[\zeta_q(2)^2 - \zeta_q(4)\bigr],\\
\frac{M_4\ga^3}{M_1}&=4\bigl[\zeta_q(2)^2 - \zeta_q(4)\bigr]
-\frac{32}{3}\bigl[\zeta_q(2)^3-3\zeta_q(2)\zeta_q(4)+2\zeta_q(6)\big].
\end{aligned}
\label{eq:Mk-ETH} 
\end{equation}

Let us consider the free energy of the ETH matrix model.
$F_g$ can be obtained by 
plugging \eqref{eq:psi-ETH}, \eqref{eq:M1-ETH}, and
\eqref{eq:Mk-ETH} into the general formula in the previous section.
For instance, the genus-zero free energy is obtained 
from \eqref{eq:F0-psi} using $\psi(b)$ in \eqref{eq:psi-ETH} 
\begin{equation}
\begin{aligned}
 F_0&=\sum_{r=1}^\infty\frac{1}{4r}q^{r^2}(q^r+q^{-r}+2)+\log\ga.
\end{aligned} 
\label{eq:F0-ETH}
\end{equation}
Similarly, $F_1$ and $F_2$ are obtained 
from \eqref{eq:F1} and \eqref{eq:F2-result} using the moments $M_k$
in  \eqref{eq:M1-ETH}, and
\eqref{eq:Mk-ETH}
\begin{equation}
\begin{aligned}
F_1&=-\qu\log(q;q)_\infty,\\
 F_2&=\frac{1}{(q;q)_\infty^6}\left[
\frac{191 \zeta_q(2)^3}{180}-\frac{17 \zeta_q(2)^2}{240}+\frac{13 \zeta_q(2)
   \zeta_q(4)}{6}+\frac{\zeta_q(2)}{40}-\frac{\zeta_q(4)}{24}+
\frac{35 \zeta_q(6)}{36}-\frac{1}{240}\right].
\end{aligned} 
\label{eq:F12-ETH}
\end{equation}
Note that the last term of $F_2$ agrees with the free energy of the Gaussian 
matrix model in \eqref{eq:Fg-gauss}
\begin{equation}
\begin{aligned}
 \frac{B_{2g}}{2g(2g-2)}\Bigg|_{g=2}=-\frac{1}{240}.
\end{aligned} 
\end{equation}

\subsubsection{Small $q$ expansion of free energy}
It is natural to ask how the ETH matrix model is deformed from
the Gaussian matrix model when we turn on $q\ne0$.
To see this, let us consider the small $q$ expansion of the free 
energy of the ETH matrix model. 
In the small $q$ limit, $F_0$ in \eqref{eq:F0-ETH}
and $F_{1,2}$ in \eqref{eq:F12-ETH} are expanded as
\begin{equation}
\begin{aligned}
 F_0&=\frac{1}{4}+\frac{q}{2}+\frac{3 q^2}{8}+\cO(q^4)+\log\ga,\\
F_1&=\frac{q}{4}+\frac{3 q^2}{8}+\frac{q^3}{3}+\cO(q^4),\\
F_2&=-\frac{1}{240}+\frac{15 q^3}{4}+\cO(q^4).
\end{aligned} 
\label{eq:Fg-smallq}
\end{equation}
As we will see below, this expansion is consistent with
the direct computation of the partition function of
the ETH matrix model
\begin{equation}
\begin{aligned}
 \cZ&=\ga^{N^2}\int dM e^{-N\Tr V_0(M)}\\
&=\ga^{N^2}e^{-N^2V_0(0)}\int dM e^{-N\Tr \bigl[V_0(M)-V_0(0)\bigr]},
\end{aligned} 
\label{eq:ZintETH}
\end{equation}
where $V_0(M)$ is given by (see \eqref{eq:V0-psi})
\begin{equation}
\begin{aligned}
 V_0(M)&=-\sum_{r=1}^\infty\frac{1}{r}\psi(2r)T_{2r}(M/2),\quad
V_0(0)=-\sum_{r=1}^\infty\frac{(-1)^r}{r}\psi(2r),
\end{aligned} 
\end{equation}
and $\psi(2r)$ is the cap amplitude in \eqref{eq:psi-ETH}.
One can easily show that the small $q$-expansion
of the
potential $V_0(M)-V_0(0)$ starts from the Gaussian term
\begin{equation}
\begin{aligned}
 V_0(M)-V_0(0)=\frac{M^2}{2}+\left(\frac{3M^2}{2}-\frac{M^4}{4}\right)q+\cO(q^2).
\end{aligned} 
\label{eq:V-V0}
\end{equation}
Then the integral in \eqref{eq:ZintETH} is expanded around the Gaussian potential
in the small $q$ limit
\begin{equation}
\begin{aligned}
 \int dM e^{-N\Tr \bigl[V_0(M)-V_0(0)\bigr]}=
\int dM e^{-\frac{N}{2}\Tr M^2}\left(1+\sum_{k=1}^\infty f_k(M)q^k\right).
\end{aligned}
\label{eq:exp-Gauss} 
\end{equation}
For instance, $f_1(M)$ is given by
\begin{equation}
\begin{aligned}
 f_1(M)=-N\Tr\left(\frac{3M^2}{2}-\frac{M^4}{4}\right).
\end{aligned} 
\end{equation}
The integral \eqref{eq:exp-Gauss} can be evaluated
order by order in the small $q$ expansion
\begin{equation}
\begin{aligned}
 \cZ&=\cZ_G\ga^{N^2}e^{-N^2V_0(0)}\left(1+\sum_{k=1}^\infty z_k q^k\right),
\end{aligned} 
\label{eq:Z-smallq}
\end{equation}
where $z_k$ is the Gaussian average of $f_k(M)$
\begin{equation}
\begin{aligned}
 z_k=\frac{1}{\cZ_G}\int dM e^{-\frac{N}{2}\Tr M^2}f_k(M).
\end{aligned} 
\end{equation}
It turns out that $z_k$ is a finite Laurent polynomial in $N$.
Thus we can determine the $N$ dependence of $z_k$ from
the first few data of $z_k$ for small $N~(N=1,2,\cdots)$.
In this way, we find
\begin{equation}
\begin{aligned}
 z_1&=\qu-N^2,\\
z_2&=\frac{13}{32}+\frac{N^2}{8}+\frac{N^4}{2},\\
z_3&=\frac{15}{4 N^2}+\frac{55}{128}-\frac{55 N^2}{48}
-\frac{N^4}{4}-\frac{N^6}{6}.
\end{aligned} 
\end{equation}
Taking the log of  \eqref{eq:Z-smallq} and collecting the terms
of $\cO(N^{2-2g})$, we can compute  $F_g$ in 
the small $q$ expansion. 
We have checked that the small $q$ expansion of
$F_g$ in \eqref{eq:Fg-smallq} is correctly reproduced from \eqref{eq:Z-smallq}.

We should stress that our $\psi(b)$ 
in \eqref{eq:psi-ETH} is different from $\psi(b)$ in 
\cite{Blommaert:2025rgw}. 
We believe that our $\psi(b)$ 
in \eqref{eq:psi-ETH} is the correct cap amplitude for the ETH matrix model 
of DSSYK.
In fact, as we have seen above, the general formula of $F_0$ in \eqref{eq:F0-psi} 
with our $\psi(b)$
is consistent with the small $q$ expansion of $\cZ$
in \eqref{eq:Z-smallq}, which serves as strong evidence for the validity
of our $\psi(b)$ in \eqref{eq:psi-ETH}. 

\section{Discussion}\label{sec:discussion}
In this paper we introduced the cap amplitude $\psi(b)$
as the expansion coefficient of the 1-form $ydx$ on the spectral curve 
(see \eqref{eq:ydx-psi}).
The dilaton equation \eqref{eq:dilaton}
for the discrete volume $N_{g,n}$ can be interpreted geometrically as 
the operation of closing one of the boundaries by gluing the cap amplitude
along the boundary. In this process, 
one of the boundaries is capped and 
the number of boundaries decreases by one:
$N_{g,n+1}\to N_{g,n}$.
Similarly,
the genus-$g$ free energy $F_g$ is obtained from 
$N_{g,1}$ by gluing  the cap amplitude (see \eqref{eq:conj}).
It turns out that the 
cap amplitude is the most basic building block of the random matrix model:
$N_{g,n}$ and $F_g$ are given by some combination of the moments $M_k,J_k$, which
in turn are constructed from the cap amplitude (see \eqref{eq:MJ-psi}).
Therefore, $N_{g,n}$ and $F_g$ are completely determined once we know the 
cap amplitude $\psi(b)$, at least in principle.

There are several interesting open questions.
As discussed in \cite{norbury2013polynomials}, 
in addition to the dilaton equation, the discrete volume satisfies 
the so-called string equation.
It would be interesting to see if the string equation has a simple
expression in terms of the cap amplitude.

In \cite{Kostov:2009nj,Kostov:2010nw}, it was shown that the
genus expansion of the matrix model can be obtained by
the CFT technique.
It would be interesting to understand the CFT interpretation of the
cap amplitude.

As an application of our formalism,
it would be interesting to study the quantum cosmology 
using the ETH matrix model for DSSYK, along the lines of \cite{Blommaert:2025rgw}. 
As discussed in \cite{Okuyama:2025hsd}, de Sitter JT gravity is obtained from
the ETH matrix model by zooming in on the upper edge of the cut.
In this setup, the Hartle-Hawking wavefunction of de Sitter JT gravity
is given by $\bra \Tr e^{\ri\ell M}\ket$
\cite{Maldacena:2019cbz}, which has a natural
decomposition
\eqref{eq:Zdisk-psi} into the cap amplitude and the trumpet with $\bt=-\ri\ell$.
Here $\ell$ is interpreted as the size of the future boundary of
the de Sitter space.
As discussed in \cite{Blommaert:2025rgw},
the picture \eqref{fig:Zdisk} suggests that the cap amplitude
has a natural interpretation as the 
Hartle-Hawking no boundary state. It would be interesting to
make this connection more explicit.

\acknowledgments
The author is grateful to Shota Komatsu for useful discussions.
This work was supported
in part by JSPS Grant-in-Aid for Transformative Research Areas (A) 
``Extreme Universe'' 21H05187 and JSPS KAKENHI 25K07300.

%%%%%%%%%%%%%%%
\bibliography{paper}
\bibliographystyle{utphys}

\end{document}